# Structural and Physical Properties of CaFe$_4$As$_3$ Single Crystals

A.B. Karki,[1] G.T. McCandless,[2] S. Stadler,[1] Y.M. Xiong,[1] J. Li,[1] Julia Y. Chan,[2] and R. Jin[1]

[1]*Department of Physics and Astronomy, Louisiana State University, Baton Rouge, Louisiana 70803, USA*

[2]*Department of Chemistry, Louisiana State University, Baton Rouge, Louisiana 70803, USA*

Abstract

We report the synthesis, and structural and physical properties of CaFe$_4$As$_3$ single crystals. Needle-like single crystals of CaFe$_4$As$_3$ were grown out of Sn flux and the compound adopts an orthorhombic structure as determined by X-ray diffraction measurements. Electrical, magnetic, and thermal properties indicate that the system undergoes two successive phase transitions occurring at $T_{N1} \sim 90$ K and $T_{N2} \sim 26$ K. At $T_{N1}$, electrical resistivities ($\rho_b$ and $\rho_{ac}$) are enhanced while magnetic susceptibilities ($\chi_b$ and $\chi_{ac}$) are reduced in both directions parallel and perpendicular to the *b*-axis, consistent with the scenario of antiferromagnetic spin-density-wave formation. At $T_{N2}$, specific heat reveals a slope change, and $\chi_{ac}$ decreases sharply but $\chi_b$ has a clear jump before it decreases again with decreasing temperature. Remarkably, both $\rho_b$ and $\rho_{ac}$ decrease sharply with thermal hysteresis, indicating the first-order nature of the phase transition at $T_{N2}$. At low temperatures, $\rho_b$ and $\rho_{ac}$ can be described by $\rho = \rho_0 + AT^\alpha$ ($\rho_0$, A, and α are constants). Interestingly, these constants vary with applied magnetic field. The ground state of CaFe$_4$As$_3$ is discussed.

The recent discovery of superconductivity with high transition temperature ($T_c$) in iron-based materials has attracted enormous attention in the condensed matter and materials physics community.[1-7] While the superconducting mechanism remains unclear, it is commonly agreed that FeAs or FeSe building blocks play a key role in the emergence of superconductivity. In general, superconductivity occurs when the structural and antiferromagnetic (AFM) spin-density-wave (SDW) transitions – which exist in the parent compounds – are suppressed upon chemical doping or the application of hydrostatic pressure.[4,6,8-13] This makes the search for new Fe-based superconductors inseparable from the search for compounds that form AFM SDWs. Recently, a new FeAs-based compound, CaFe$_4$As$_3$, has been discovered, in which FeAs$_4$ tetrahedra form a ribbons along the *b* direction and a rectangular network in the *ac* plane. More prominent is that the system undergoes two successive AFM SDW transitions,[14,15] with one incommensurate (IC-SDW) at $T_{N1} \sim 90$ K along the *b* axis and the other commensurate (C-SDW) at $T_{N2} \sim 26$ K in the *ac* plane.[16,17] However, its other physical properties are much less clear. In this article, we report the crystal growth, and structural and physical properties of $CaFe_4As_3$ including electrical and thermal conductivity, Hall effect, magnetic susceptibility, specific heat, and thermopower.

Single crystals of $CaFe_4As_3$ were grown out of Sn flux. The stoichiometric amounts of high purity calcium shot (99.999 % Alfa Aesar), iron powder (99.95 % Alfa Aesar), and arsenic powder (99.999 % Alfa Aesar) were mixed with tin shot (99.999 % Alfa Aesar) in the ratio 1:4:3:40.[14] The mixture was placed in an alumina crucible and sealed in a quartz tube under vacuum. The alumina crucible was plugged lightly with quartz wool before the sealing. The whole assembly was heated in a box furnace to 600 °C at a rate of 200 °C/h and was held at this temperature for 5 h. It was further heated to 1100 °C at a rate of 200 °C/h where it was held for 15 h, and then cooled to 600 °C at a rate of 10 °C/h. At this temperature, the molten Sn was decanted by spinning the tube in a centrifuge. Shiny needle shaped single crystals of rectangular cross-section were obtained with a typical size of 0.2 x 0.2 x 5 $mm^3$. The as-grown crystals were etched with dilute nitric acid to remove the residual Sn flux. Fig. 1a shows the needle-like single crystals.

The crystal structure was determined by single-crystal X-ray diffraction. Single-crystal X-ray diffraction is performed using Nonius KappaCCD X-ray diffractometer equipped with a Mo $K_\alpha$ radiation source ($\lambda = 0.71073$ Å) and graphite monochromator. The crystal used for this experiment was selected from the same batch as that used for physical properties measurements and mounted on a glass fiber with epoxy. The size of this crystal was approximately 0.025 x 0.075 x 0.125 $mm^3$. The refinement results confirm that the crystal is indeed $CaFe_4As_3$ with an orthorhombic space group symmetry *Pnma*. The lattice parameters are $a = 11.918(2)$ Å, $b = 3.749(1)$ Å, and $c = 11.625(3)$ Å. The calculated density of $CaFe_4As_3$ is 6.24 $g/cm^3$. A schematic view of the crystal structure of $CaFe_4As_3$ is shown in Fig. 1b. The parallel FeAs ribbons are aligned along the *b* axis while forming rectangular network in the *ac* plane. The phase purity is further confirmed by powder X-ray diffraction which is carried out by a Scintag XDS2000 powder X-ray diffractometer using Cu $K_\alpha$ radiation ($\lambda = 1.54056$ Å). Fig. 1c shows the X-ray powder diffraction pattern of $CaFe_4As_3$. Note that the sample is of single phase.

Electrical resistivity of $CaFe_4As_3$ was measured using a standard four-probe A.C. technique at 27 Hz with an excitation current of 1 mA. Thin Pt wires were attached to the sample using a conductive epoxy (Epotek H20E). Data was collected in the temperature range of 2 to 300 K and in magnetic fields up to 14 T using a *Quantum Design* Physical Property Measurement System (PPMS). Fig. 2a shows the temperature dependence of electrical resistivities ($\rho_b$ and $\rho_{ac}$) in both cooling and warming modes, where $\rho_b$ and $\rho_{ac}$ are resistivities parallel and perpendicular to the *b* axis, respectively. Note that both $\rho_b$ and $\rho_{ac}$ exhibit a similar temperature dependence with $\rho_{ac}/\rho_b(290K) \sim 1.3$, even though the parameters *a* and *c* are much longer than *b*. This is comparable to that of other FeAs-based layered compounds. Nevertheless, similar to previous reports,[14,15] both $\rho_b$ and $\rho_{ac}$ decrease as the temperature is lowered from room temperature to $T_{N1}$ (~ 90 K). Below $T_{N1}$, both $\rho_b$ and $\rho_{ac}$ deviate from their high-temperature behavior with relatively enhanced magnitude (see Figs. 2b-2c). As temperature is further lowered, we note that a sharp drop occurs in both $\rho_b$ and $\rho_{ac}$ at $T_{N2} \sim 26$ K. According to neutron scattering measurements,[16,17] the system forms an incommensurate spin density wave (IC-SDW) along the *b* axis below $T_{N1}$, but a commensurate spin density wave (C-SDW) in the *ac* plane below $T_{N2}$. Thus, the anomalies seen in both $\rho_b$ and $\rho_{ac}$ must be associated with these two transitions. However, the impact on the electrical transport due to these two transitions are different: there are small changes in both $\rho_b$ and $\rho_{ac}$ at $T_{N1}$ without any thermal hysteresis (Figs. 2b-2c) but have sharp decrease at $T_{N2}$ with obvious thermal hysteresis as illustrated in Figs. 2d-2e. This strongly suggests that a second-order transition at $T_{N1}$, where as the transition at $T_{N2}$ is first order.

The formation of a SDW usually involves the opening of an energy gap at the Fermi level, which should result in enhanced electrical resistivity. The resistivity anomaly at $T_{N1}$ is consistent with this picture. More unconventional is the sharp resistivity drop at $T_{N2}$. This is likely due to the reduction of spin scattering with the formation of a C-SDW. It should be pointed out anomalies in both $\rho_b$ and $\rho_{ac}$ are reported for the first time, where as previous reports show either a change in $\rho_{ac}$[14] or $\rho_b$[15] at $T_{N2}$ only. We note that $\rho_b$ measured in crystals with large cross sectional areas is low with the absence of anomalies at both $T_{N1}$ and $T_{N2}$. Such samples are found to consist of multiple single crystals stacked in parallel with trapped Sn flux. This provides low-resistivity paths for the current along the $b$ axis and leads to sharp resistivity drop around 3.8 K (~ $T_c$ of Sn). The impact of Sn impurity on $\rho_{ac}$ is, on the other hand, negligible as the current is normal to the Sn paths. Nevertheless, the application of magnetic field seems to push $T_{N2}$ to higher temperatures, as demonstrated in Figs. 3a – 3c, while $T_{N1}$ remains unchanged for fields up to 14 Tesla (not shown here). This strongly suggests that magnetic field promotes the transition from longitudinal IC-SDW to transverse C-SDW. The question is whether the same group of electrons is involved in both transitions. To answer this, we performed Hall effect measurements by applying $H \perp I // b$. Shown in Fig. 4 is the temperature dependence of Hall coefficient $R_H$. Note that there are corresponding changes at both $T_{N1}$ and $T_{N2}$. In particular, there is sharp increase in the magnitude of $R_H$ at $T_{N2}$. While we cannot determine the change of carrier concentration due to its multi-band nature,[17] it is conceivable to state that the carrier concentration decreases below $T_{N2}$. The continuous variation of $R_H$ with the sign change at $T_{N2} < T < T_{N1}$ indicates the continuous transition (second-order nature) below $T_{N1}$.

Nevertheless, the finite electrical resistivity at low temperatures indicates that the Fermi surface is only partially gapped and the remaining electrons have a metallic ground state ($d\rho_i/dT > 0$, $i = b, ac$). Quantitatively, both $\rho_b$ and $\rho_{ac}$ can be fitted to a power law,

$$\rho = \rho_0 + AT^\alpha, \quad\quad (1)$$

in the temperature range between 2 and 15 K, where $\rho_0$, A, and $\alpha$ are $T$-independent constants. As shown in Figs. 5a – 5b, Eq. (1) describes our experimental data very well for both $\rho_b(T)$ and $\rho_{ac}(T)$ below 15 K (solid curves), with $\alpha = 1.76$ and 1.83 at $H = 0$, respectively. In contrast to previous reports,[15] our result suggests that the ground state of CaFe$_4$As$_3$ is non-Fermi-liquid like, since both $\rho_b(T)$ and $\rho_{ac}(T)$ deviate from $T^2$ dependence. Remarkably, the $\alpha$ value for $\rho_b(T)$ increases with increasing $H$ (Fig. 5c). Note that $\alpha$ for $\rho_b(T)$ approaches 2 at $H = 14$ T, implying the recovery of Fermi-liquid behavior at high fields along the $b$ axis. On the other hand, $\alpha$ for $\rho_{ac}(T)$ is more or less field independent. This suggests that the C-SDW–that propagates in the $ac$ plane–is robust under applied magnetic fields. Interestingly, the residual resistivity $\rho_0$ increases with field in both the $b$ axis and $ac$ plane (Fig. 5d). This is unexpected if it is simply resulted from non-magnetic impurities/defects. The fact that both $\rho_0$ and A are field dependent (see Figs. 5d-5e) most likely reflects strong magnetic interactions in the remaining electronic system, even though they do not form SDWs.

The magnetoresistance $MR$ [$MR = \{\rho(H)-\rho(H=0)\}/\rho(H=0)$] of CaFe$_4$As$_3$ below $T_{N2}$ makes it more clear. Figs. 3d - 3e display the field dependence of both transverse ($H \perp I // b$) and longitudinal ($H // I // b$) $MR$ at $T < T_{N2}$, respectively. Several features are worth noting: (1) both transverse and longitudinal $MR_b$ are positive below $T_{N2}$, (2) the transverse $MR_b^\perp$ is much larger than the longitudinal one ($MR_b^{//}$), and (3) $MR_b^\perp$ exhibits linear field dependence in a wide field range below $T_{N2}$. In addition, our data do not fall on a single line when plotted as $MR_b^\perp$ versus $H/\rho_0$ (not shown), indicating a violation of Kohler's rule. This is likely the consequence of the non-Fermi liquid ground state as discussed earlier. Similar linear field dependence of magnetoresistivity has been observed in other FeAs-based compounds such as CaFe$_2$As$_2$.[12] While the origin is unknown, this cannot be attributed to the Hall contribution, since our sample is long and thin. The fact that $MR_b^\perp > MR_b^{//}$ indicates there is strong spin-orbital coupling. Nevertheless, the positive $MR$ along both the $b$ and $ac$ directions suggests AFM type magnetic interactions. The application of magnetic field tends to break such interactions leading to enhanced spin scattering, even though the effect along the $ac$ plane is much smaller (see Fig. 3f).

There were previously two reports in magnetic susceptibility anisotropy: one shows $\chi_b > \chi_{ac}$ [15] and the other $\chi_b < \chi_{ac}$ [14] above $T_{N1}$. We measured the magnetic susceptibility anisotropy of CaFe$_4$As$_3$ with 10 mg sample using a Superconducting Quantum Interference Device (SQUID) magnetometer (*Quantum Design*), from $T = 2$ to 400 K and with a vibrating sample magnetometer (VSM) from 300 to 700 K. Fig. 6a displays the temperature dependence of $\chi_b$ and $\chi_{ac}$ between 2 and 700 K. Note that there is little anisotropy between $\chi_b$ and $\chi_{ac}$, both decreasing with increasing temperature above $T_{N1}$. Above ~ 400 K, the susceptibility follows a Curie-Weiss behavior as demonstrated in the inset of Fig. 6a. The solid line is the fitting curve $\chi(T) = C / (T-\theta_{cw})$ in the temperature range of 400 – 700 K with $C = 7.87$ K cm$^3$/mol and Cuire-Weiss temperature $\theta_{CW} = -632$ K. This allows us to estimate the effective magnetic moment $\mu_{eff} = 2.14$ $\mu_B$/Fe, which is comparable to that obtained from neutron scattering measurements.[17] The negative $\theta_{CW}$ indicates antiferromagnetic interaction, consistent with the scenario that system forms AFM SDW at low temperatures. Interestingly, $|\theta_{CW}| \gg T_{N1}$, which is likely the consequence of low-dimensional magnetism. Nevertheless, the temperature dependence of $\chi_b$ and $\chi_{ac}$ below 150 K is highlighted in Fig. 6b. At $T_{N1} = 90$ K, both $\chi_b$ and $\chi_{ac}$ decrease with temperature until it reaches $T_{N2}$. This is consistent with the scenario that the system forms a AFM SDW below $T_{N1}$. Given that $\chi_b$ decreases much more steeply than $\chi_{ac}$, the SDW should be along the $b$ direction, as confirmed by neutron scattering measurements.[17] Apparently, $\chi_b$ and $\chi_{ac}$ behave differently at $T_{N2} = 26$ K. While $\chi_b$ first increases before it decreases again, $\chi_{ac}$ decreases sharply down to 4 K. Again, the behaviors of $\chi_b$ and $\chi_{ac}$ below $T_{N2}$ support the scenario that the system switches from a longitudinal SDW to a transverse SDW (the easy axis is in $ac$ plane below $T_{N2}$) as determined by neutron scattering measurements.[17] The isothermal magnetization versus magnetic field data at 5 K, presented in Fig. 6c, suggests that there is a weak FM interaction along the $b$ direction in addition to the AFM interaction mentioned above. There is a small upturn in both $\chi_b$ and $\chi_{ac}$ below ~ 4 K. This should be due to the paramagnetic contribution from the remaining electrons that do not participate the SDW formation.

Based on the above analysis, it is expected that specific heat of CaFe$_4$As$_3$ would exhibit features associated with the second-order phase transition at $T_{N1}$ and the first-order transition with a latent heat at $T_{N2}$. Fig. 7a shows specific heat $C_p$ measured in the PPMS using the relaxation method between 2 and 150 K. Similar to previous report,[15] a peak was observed in specific heat at $T_{N1}$, confirming the second-order nature of the phase transition. However, $C_p(T)$ varies smoothly when crossing $T_{N2}$, and no obvious thermal hysteresis was observed when measurement was conducted via both cooling and warming. As shown in the inset of Fig. 7a, there is slope change at $T_{N2}$ when the data is plotted as $C_p/T$ vs. $T^2$. This suggests that the change at $T_{N2}$ is electronic, i.e., change of electronic specific heat coefficient, in origin and entropy due to spin fluctuation is largely removed prior to the transition.

In order to obtain details of the electronic properties, we measured the low-temperature specific heat under the application of magnetic fields up to 14 Tesla ($H \perp b$). Fig. 7b displays the data plotted as $C_p/T$ vs. $T^2$ from 2 to 13 K at various applied magnetic fields. For a non-magnetic system, the low-temperature $C_p/T$ is expected to vary linearly with $T^2$, i.e., $C_p(T)/T = \gamma + \beta T^2$ ($\gamma$ and $\beta$ are $T$-independent coefficients). These two coefficients represent the electron and phonon contributions, respectively. For CaFe$_4$As$_3$, we note that $C_p/T$ vs. $T^2$ at zero and low fields are not perfectly linear, having an upturn at low temperatures. Since it is suppressed by magnetic fields, such an upturn can be attributed to the Schottky effect. We thus fit the low-temperature $C_p/T$ data using

$$C_p/T = \gamma + \beta T^2 + \alpha' [e^{\Delta/T} / T^3 (1+e^{\Delta/T})^2] \qquad (2)$$

with the third term describing the Schottky contribution. Here, $\alpha'$ is a constant proportional to number of two-level systems, and $\Delta$ is the energy gap between the two levels. The fitting results in $\gamma = 82$ mJ mol$^{-1}$K$^{-2}$, $\beta = 1.05$ mJ mol$^{-1}$K$^{-4}$, $\alpha' = 11500$ mJ mol$^{-1}$K, and $\Delta = 6$ K for zero field. These values of $\gamma$ and $\beta$ are very close to those reported previously.[15] One can estimate the Debye temperature $\Theta_D \sim 245$ K via $\beta = N(12/5)\pi^4 R \Theta_D^{-3}$, where $R = 8.314$ J mol$^{-1}$ K$^{-1}$ and the number of atoms in one unit cell is $N = 8$ for CaFe$_4$As$_3$. This value is slightly higher than that from the theoretical calculation.[14]

Remarkably, the application of magnetic field clearly shifts the low-temperature $C_p/T(T^2)$ curve upward, indicating the change of $\gamma$, $\beta$, $\alpha'$ and $\Delta$. In particular, there is an increase in $\gamma$ with $H$ as demonstrated in the inset of Fig. 7b. This reflects the modification of electron-electron interactions by magnetic field, which leads to a change in the electronic density of states at the Fermi level $N(E_F)$ as $\gamma = \pi^2/3 \, k_B^2 N(E_F)$.[18] Recalling that the application of magnetic field results in the $T^2$ dependence of $\rho_b$, we may further calculate the Kadowaki-Woods ratio, $A/\gamma^2$, at high fields. For $H = 14$ T, $\gamma = 26$ mJ mol$^{-1}$ Fe$^{-1}$ K$^{-2}$ and $A_b = 0.45$ $\mu\Omega$ cm K$^{-2}$, which leads to $A/\gamma^2 \sim 66 \times 10^{-5}$ $\mu\Omega$ cm mol$^2$ K$^2$ mJ$^{-2}$. This value is much higher than that for free-electron system ($\sim 1 \times 10^{-6}$ $\mu\Omega$ cm mol$^2$ K$^2$ mJ$^{-2}$), suggesting that CaFe$_4$As$_3$ is a strongly correlated system.[19-22]

In magnetic systems, thermal transport properties often exhibit unusual behavior due to magnetic contributions. For the first time, we report both thermal conductivity ($\kappa_b$) and Seebeck coefficient ($S_b$) of single crystal CaFe$_4$As$_3$ along the $b$ axis measured by using the standard four-probe method. As shown in Fig. 8, $\kappa_b$ increases with increasing $T$ above $T_{N1}$. In this temperature range, the thermal conductivity is usually dominated by phonon contribution, which increases with decreasing temperature. The apparent opposite trend may result from (1) large electronic contribution, or (2) surface heat emission due to the relatively large surface area compared to its cross section which is not corrected during the measurement, or (3) magnetic contribution. We may estimate the electronic contribution via Wiedemann-Franz law $\kappa_e = 2.44 \times 10^{-8} T/\rho_b$, which is presented in Fig. 8 as well. Note that $\kappa_e$ is much smaller than $\kappa_b$, thus ruling out the first scenario. Nevertheless, $\kappa_b$ is enhanced below $T_{N1}$, indicating the suppression of spin scattering below $T_{N1}$. Unfortunately, there is insufficient information below $T_{N2}$ due to small signal/noise ratio.

On the other hand, thermopower ($S_b$) along the $b$ axis, sharply decreases and changes sign from positive above $T_{N2}$ to negative below $T_{N2}$. The result is intriguing because thermopower usually possesses the same sign as that of Hall coefficient. The sign difference between $S$ and $R_H$ could be due to either diffusion scattering,[23,24] or phonon drag effect,[25,26] or a complex Fermi surface.[23] For example, it is well-known that $YBa_2Cu_3O_7$ has a single band and is hole-like with positive $R_H$. However, the measured $S_{ab}$ is negative[23] because the minority carrier (electron) has a heavier mass than holes, thus contributing strongly to the diffusion coefficient ($A_D$) (note $S = A_D\rho$). $CaFe_4As_3$ is a multiple band system involving both electrons and holes as reflected in the sign change of $R_H$. The sign difference in thermopower may be caused by the same origin as that in $YBa_2Cu_3O_7$.[23]

In summary, we have successfully grown needle-like $CaFe_4As_3$ single crystals and investigated their structural and physical properties. Electrical and thermal conductivity, Hall effect, magnetic susceptibility, specific heat, and thermopower exhibit anomalies at $T_{N1} \sim 90$ K and $T_{N2} \sim 26$ K, which were previously identified as AFM IC-SDW and C-SDW transitions, respectively. Our investigation indicates that the transition at $T_{N1}$ is second order, and that at $T_{N2}$ is first order. The application of magnetic field tends to push $T_{N2}$ to higher temperatures, but has no obvious impact on $T_{N1}$. The apparent metallic behavior at low temperatures indicates that the Fermi surface is partially gapped by the formation of SDWs. The remaining electrons exhibit non-Fermi-liquid character in zero field but resumes Fermi-liquid character in high fields. At $H = 14$ Tesla, we estimate Kadowaki-Woods ratio $A/\gamma^2 \sim 66 \times 10^{-5}$ $\mu\Omega$ cm mol$^2$ K$^2$ mJ$^{-2}$, suggesting that $CaFe_4As_3$ is a strongly correlated system.

We acknowledge the support of NSF under Grants No. DMR-1002622 (R. J.), DMR-0545728 (S.S.), and DMR 1063735 (J. Y. C.).

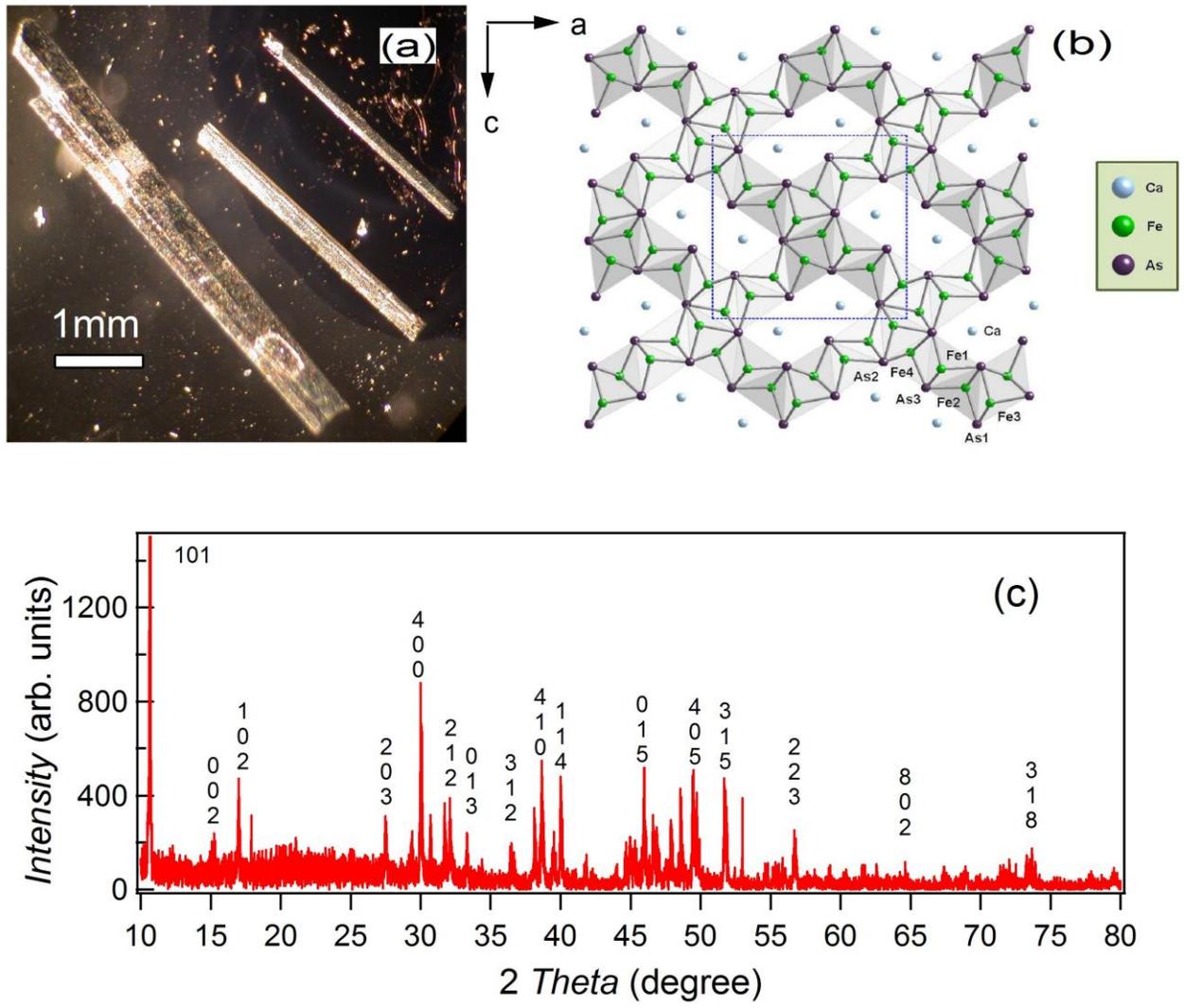

Fig. 1. (a) Picture of single crystals as grown by flux method. (b) Structure of $CaFe_4As_3$ (c) Powder X-ray diffraction pattern with indexed peaks.

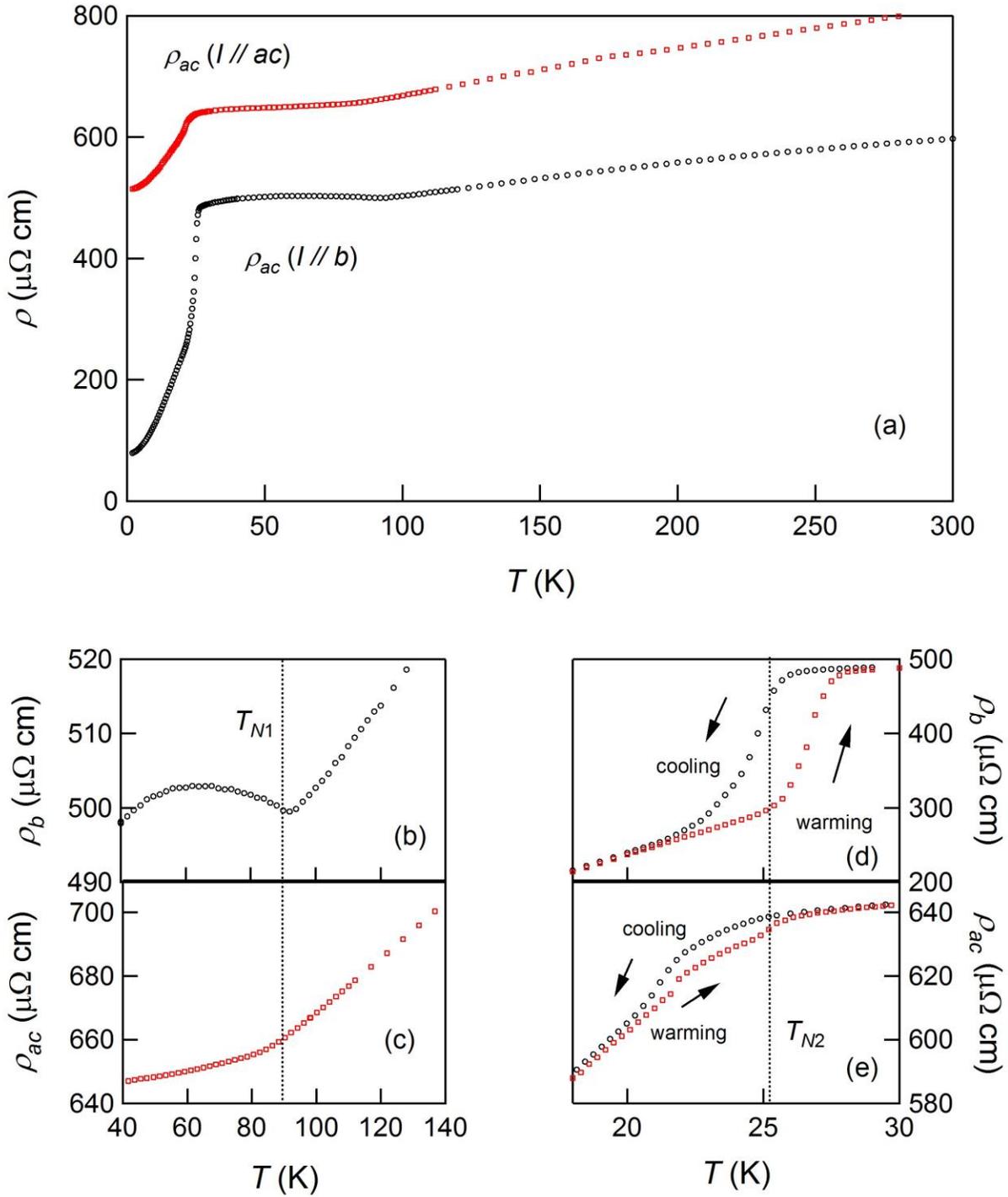

Fig. 2. (a) Temperature dependence of electrical resistivity $\rho_b$ and $\rho_{ac}$ of CaFe$_4$As$_3$ along the $b$ axis and the $ac$ plane, respectively. $\rho_b$ and $\rho_{ac}$ near $T_{N1}$ ~ 90 K are highlighted in (b) and (c), respectively. (d) and (e) show thermal hysteresis in $\rho_b$ and $\rho_{ac}$ around $T_{N2}$ ~ 26 K, respectively.

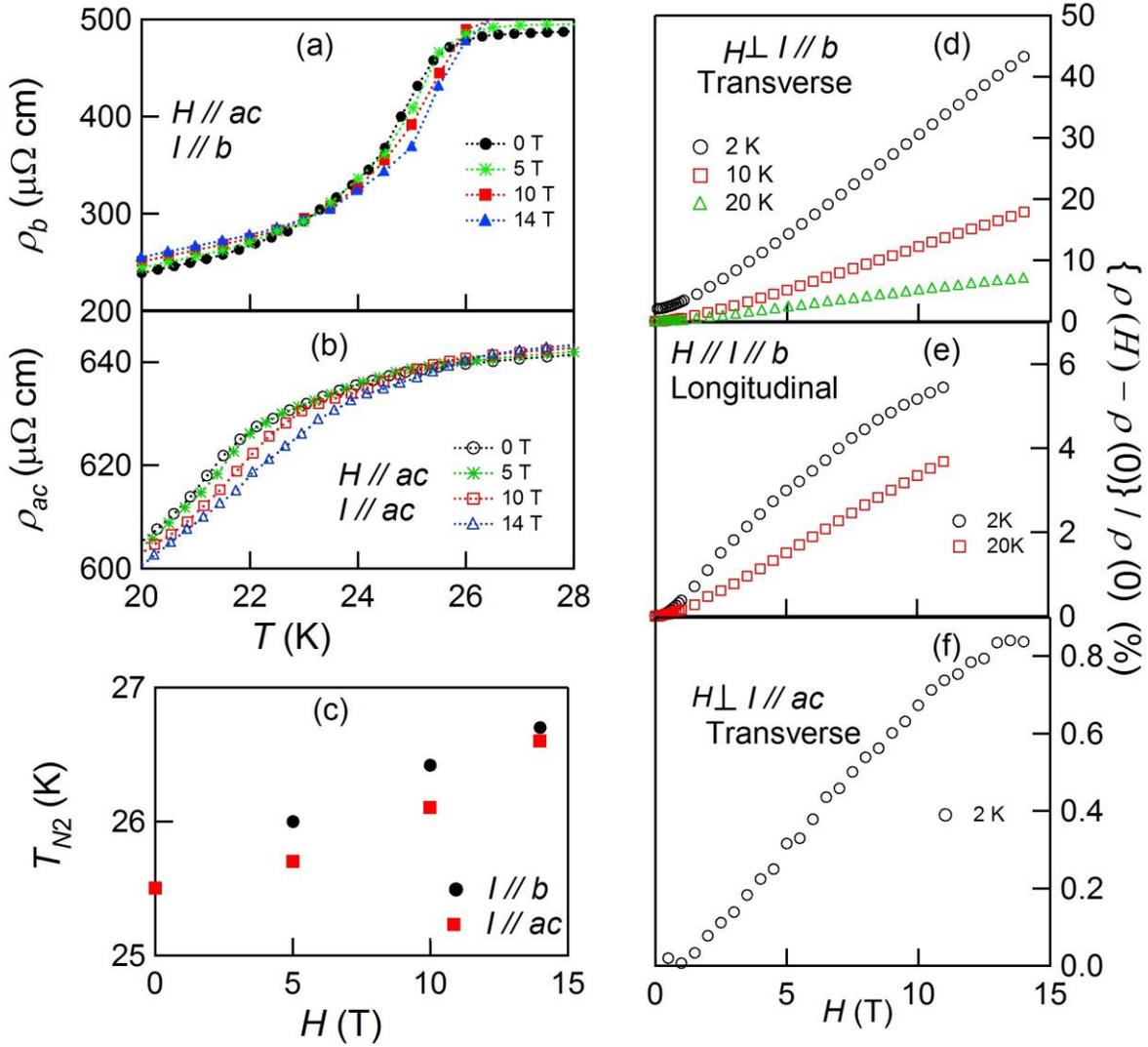

Fig. 3: Temperature dependence of $\rho_b$ (a) and $\rho_{ac}$ (b) of CaFe$_4$As$_3$ between 20 and 28 K at indicated applied magnetic fields. The dotted lines are guides to eyes. (c) Variation of $T_{N2}$ with the applied magnetic field, which is obtained at points where $\rho_b$ and $\rho_{ac}$ drop 10% compared to that extended from high-temperature resistivity. (d) Transverse magnetoresistivity along the $b$ axis ($H \perp I // b$), (e) longitudinal magnetoresistivity along the $b$ axis ($H // I // b$) (e), and (f) transverse magnetoresistivity in the ac plane ($H \perp I // ac$).

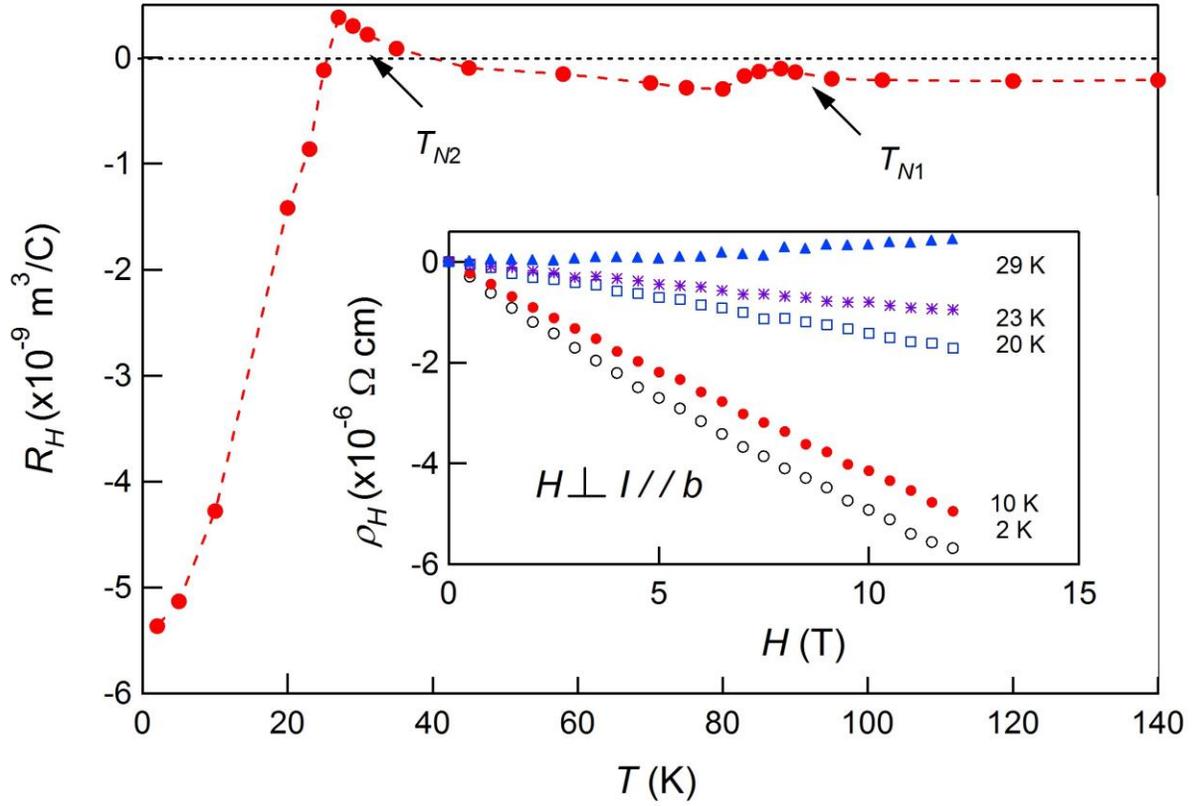

Fig. 4. Hall coefficient $R_H(T)$ for a single crystal of $CaFe_4As_3$ with anomalies at $TN_1$ and $TN_2$. The dashed line is a guide to eyes. The horizontal dotted line is a reference to $R_H = 0$. Inset: magnetic field dependence of Hall resistivity at indicated temperatures. The Hall coefficient was derived using the relation $R_H = \rho_H/H$.

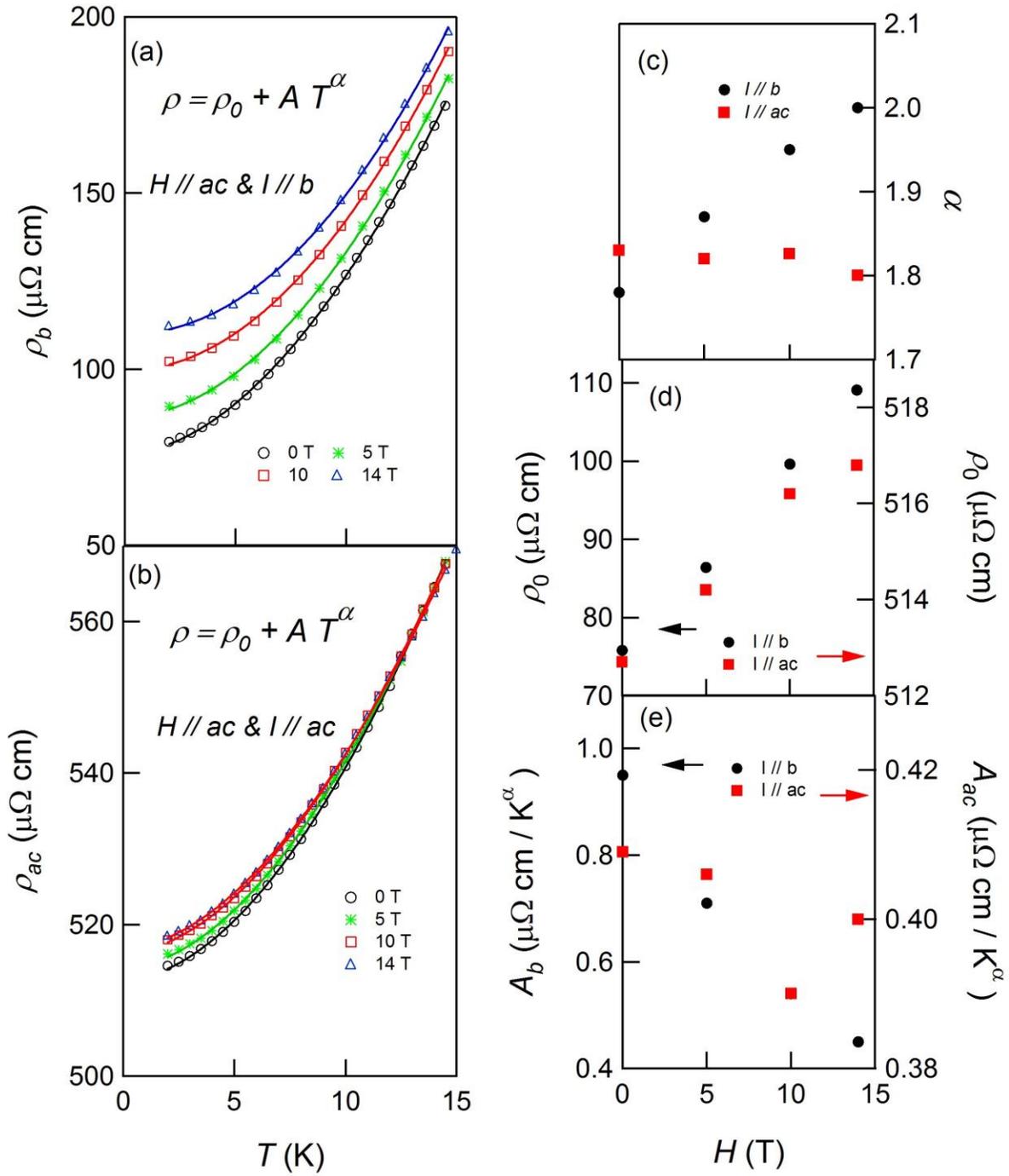

Fig. 5. Temperature dependence of $\rho_b$ (a) and $\rho_{ac}$ (b) of CaFe$_4$As$_3$ in the range of 2 to 15 K at indicated magnetic fields. The solid lines are fits to the Eq. (1). (c) - (e) show magnetic field dependence of α, $\rho_0$, and A, respectively.

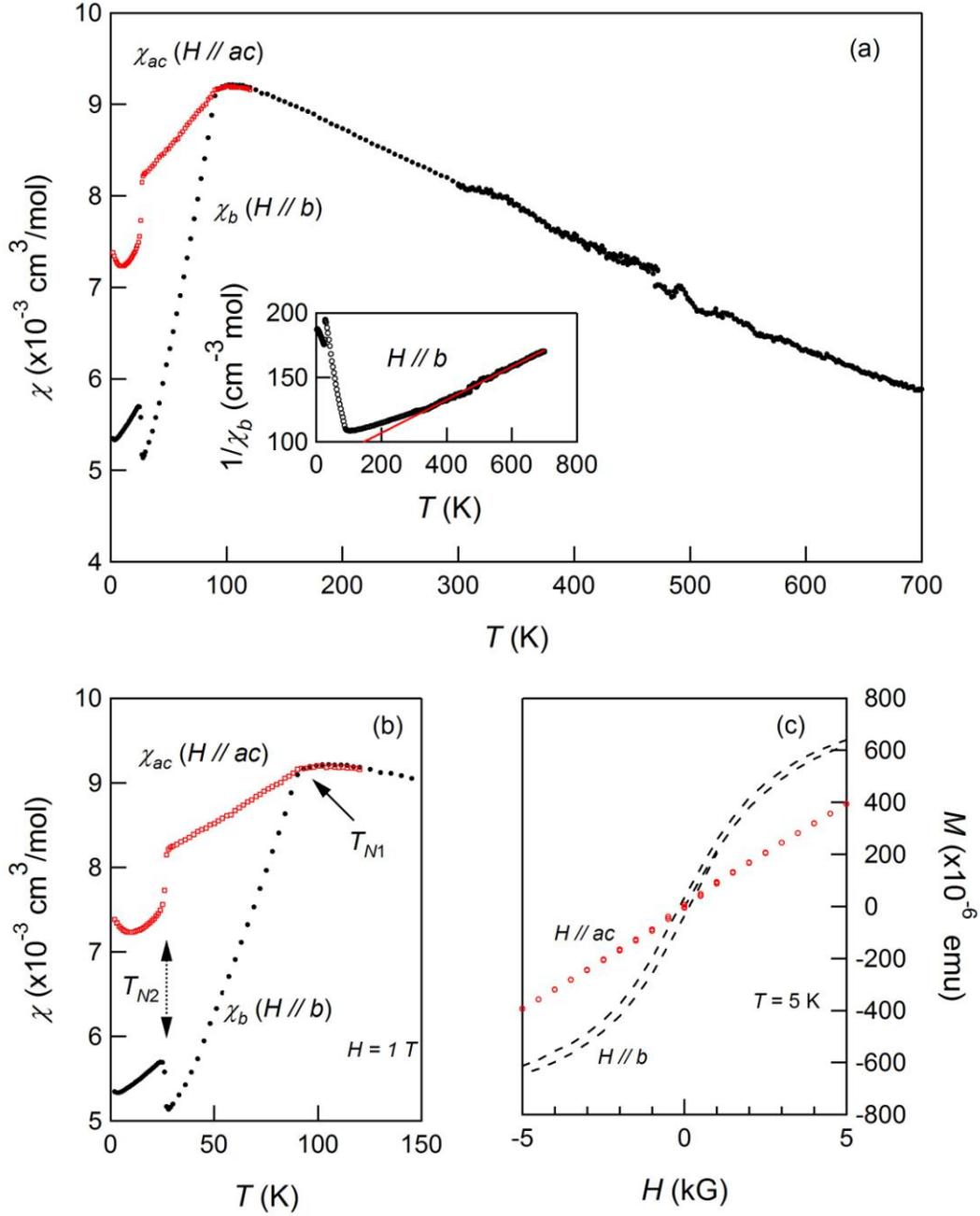

Fig. 6. (a) Temperature dependence of magnetic susceptibilities, $\chi_b(T)$ with $H // b$, and $\chi_{ac}(T)$ with $H // ac$ for CaFe$_4$As$_3$ denoted by ● and □ (color online), respectively. Data between 2 and 400 K were taken at 1 Tesla in a SQUID magnetometer and taken at $H = 10$ T between 300 to 700 K using a VSM. Inset: Temperature dependence of $1/\chi_b$ in the range of $T = 2$ to 700 K. The solid line is a fit to $\chi(T) = C / (T-\theta_{cw})$ between 400 and 700 K. (b) Magnetic susceptibility $\chi_b(T)$ and $\chi_{ac}(T)$ for $T = 2$ to 150 K at $H = 1$ T with AFM SDW transitions at $T_{N1} \sim 90$ K and $T_{N2} \sim 26$ K. (c) $M$ vs. $H$ for CaFe$_4$As$_3$ at 5 K.

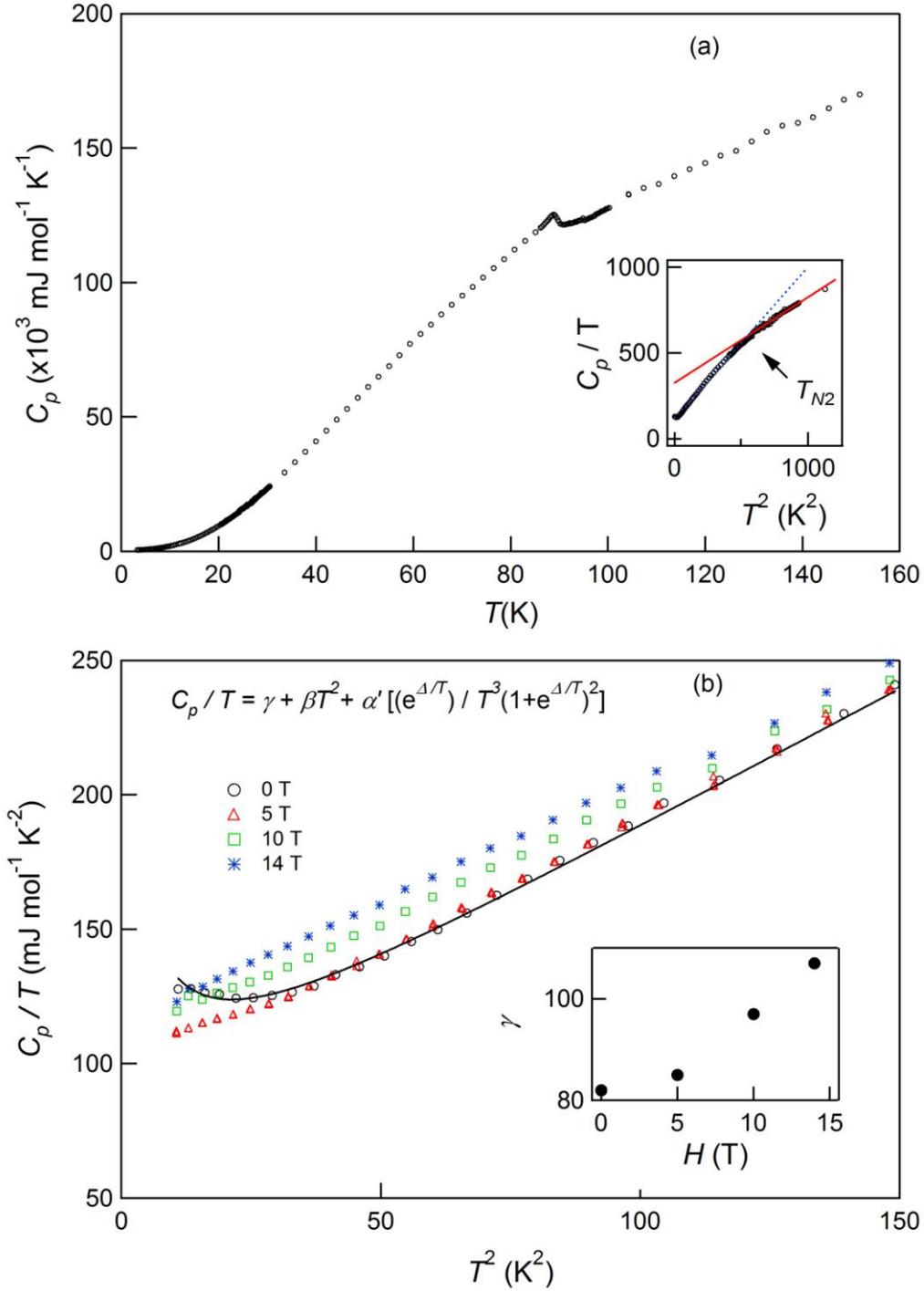

Fig.7. (a) Temperature dependence of specific heat $C_p$ of CaFe$_4$As$_3$. Inset: The change in the slope of $C_p/T$ vs. $T^2$ at $T_{N2}$. (b) $C_p/T$ vs. $T^2$ in the range of 2 to 15 K at indicated magnetic fields. The solid line is the fit of zero-field $C_p/T$ to Eq. (2). Inset: variation of $\gamma$ with $H$.

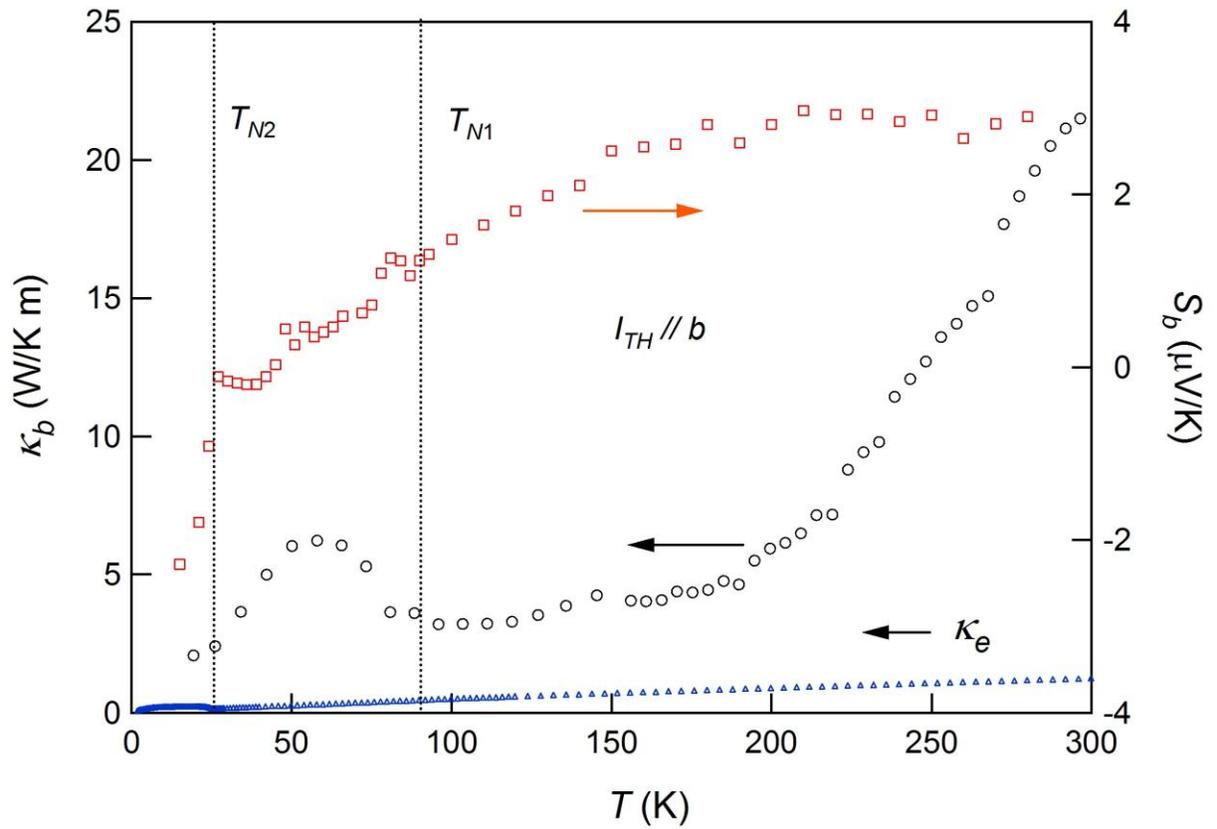

Fig. 8. Thermal conductivity (left axis) and Seebeck coefficient (right axis) of a CaFe$_4$As$_3$ single crystal for $T$ = 15 to 300 K. For comparison, the estimated electronic thermal conductivity $\kappa_e$ is also plotted.